\documentclass[article,a4paper]{revtex4-2}

\usepackage{amsmath,amssymb,amsthm,amsfonts,mathrsfs,bm,verbatim}
\usepackage{graphicx,subfigure}

\usepackage[justification=centering]{caption}
\usepackage[colorlinks=true, allcolors=blue]{hyperref}

\def\be{\begin{equation}}
\def\ee{\end{equation}}
\def\ba{\begin{eqnarray}}\def\bea{\begin{eqnarray}}
\def\ea{\end{eqnarray}}  \def\eea{\end{eqnarray}}

\def\l{\lambda}

\begin{document}
\title{Cosmological horizons from classical double copy}
\author{Jun-Lang He$^{1,2}$}\author{Jia-Hui Huang$^{3,4}$\footnote{huangjh@m.scnu.edu.cn}}
\affiliation{$^{1}$Key Laboratory of Atomic and Subatomic Structure and Quantum Control (MOE), Guangdong Basic Research Center of Excellence for Structure and Fundamental Interactions of Matter, Institute of Quantum Matter, South China Normal University, Guangzhou 510006, China}
\affiliation{$^{2}$Guangdong-Hong Kong Joint Laboratory of Quantum Matter, Guangdong Provincial Key Laboratory of Nuclear Science, Southern Nuclear Science Computing Center, South China Normal University, Guangzhou 510006, China}
\affiliation {$^{3}$Key Laboratory of Atomic and Subatomic Structure and Quantum Control (Ministry of Education), Guangdong Basic Research Center of Excellence for Structure and Fundamental Interactions of Matter, School of Physics, South China Normal University, Guangzhou 510006, China}
\affiliation {$^{4}$Guangdong Provincial Key Laboratory of Quantum Engineering and Quantum Materials, Guangdong-Hong Kong Joint Laboratory of Quantum Matter, South China Normal University, Guangzhou 510006, China}

\begin{abstract}
The classical doubly copy provides relations between classical solutions in gravitational theories and solutions in gauge theories. In this paper, we consider the cosmological horizons in the gravity side from the perspective of gauge theories in the classical double copy paradigm. We give several examples to show how to locate the cosmological horizons
by using only the single- and zeroth-copy data on the base spacetime. It is found that the proposed double copy procedure holds not only for cases with flat base spacetime but also 
for certain case with curved base spacetime.    
\end{abstract}
\maketitle

\section{Introduction}
\label{sec:1}

The relationships between gauge and gravity theories have been actively explored for decades. A remarkable discovery is the perturbative double copy \cite{Bern:2008qj,Bern:2010yg,Bern:2010ue}, which relates scattering amplitudes of the gauge and gravity theories. The success of the perturbative double copy has motivated investigations into classical solutions. Surprisingly, it turns out that some exact classical solutions of Einstein equations can be obtained from classical solutions of Abelian Yang-Mills (Maxwell) equations and associated auxiliary scalar fields, which is the classical doubly copy. Conventionally, the gravitational solution is called the ``double copy'', while the gauge solution is called the ``single copy'' of the gravitational solution and the scalar is termed the ``zeroth copy''.

Kerr-Schild double copy is a notable framework of the classical double copy, which was first proposed in \cite{Monteiro:2014cda}. 
Spacetime metric which admits Kerr-Schild coordinates can be written as an obviously double copy form of a Maxwell gauge field 
(accurately, an additional Klein-Gordon scalar field is also needed). A typical example is that the Schwarzschild metric sourced by a point mass at the origin is related to the gauge field sourced by a point charge at the origin. After the pioneer work \cite{Monteiro:2014cda}, the Kerr-Schild double copy forms of a number of Kerr-Schild spacetime solutions have 
been studied in literature 
 \cite{Luna:2015paa,Luna:2016due,Carrillo-Gonzalez:2017iyj,Ilderton:2018lsf,Berman:2018hwd,CarrilloGonzalez:2019gof,Pasarin:2020qoa,
 Easson:2020esh,Alkac:2021bav,Elor:2020nqe,Gumus:2020hbb,Alkac:2021seh,Alkac:2022tvc,Spallucci:2022xtr,Dempsey:2022sls,Easson:2023dbk}. By the way, there is another framework of classical double copy for classical solutions of gravity and gauge fields which is called Weyl double copy that relates the field strengths of the two sides \cite{Luna:2018dpt,White:2020sfn,Godazgar:2020zbv,Chacon:2021wbr,Monteiro:2020plf,Chacon:2021lox,Easson:2021asd,Han:2022ubu,
Han:2022mze,CarrilloGonzalez:2022mxx,Luna:2022dxo,Easson:2022zoh,CarrilloGonzalez:2022ggn,Armstrong-Williams:2023ssz,Alkac:2023glx}.

The main idea of the double copy paradigm is to understanding gravity theory through gauge theory, i.e. quantities in gravity theory are expected to be described by quantities in the corresponding gauge theory \cite{Adamo:2022dcm}. Kerr-Schild double copy of classical solutions responds the idea from the aspect of the construction of local fields. 
Black holes are important and interesting classical solutions in the gravity side and horizons are typical structures of black holes. 
In \cite{Easson:2020esh}, the authors found an interesting 
connection between black hole horizons and single copy gauge field: when the spacetime has multiple horizons, the single-copy electric field changes signs between horizon radii. Recently, a procedure for locating black hole horizons in Kerr-Schild spacetimes using only single- and zeroth-copy data on flat base spacetime is given in \cite{Chawla:2023bsu}.
Motivated by these works, we here address the problem of locating the cosmological horizons in the gravity side by using just associated single- and zeroth-copy data on the base spacetime. 

This paper is organised as follows. Section II is a brief review of the Kerr-Schild double copy. In Section III, we introduce the local definition of cosmological horizon of a spacetime metric which admits a Kerr-Schild form. The quantities that play important roles in our discussion are expansions of null geodesic congruences.  In Section IV, we calculate several examples to show how to use the single- and zeroth-copy data on base spacetime to determine the corresponding cosmological horizons. The final section is devoted to a summary.

\section{Kerr-Schild double copy}

Here we give a brief review of the Kerr-Schild double copy, first introduced in \cite{Monteiro:2014cda}. Considers a spacetime metric that admits a Kerr-Schild form with a flat base metric, i.e. 
\begin{equation}\label{eq2.1.1}
    g_{\mu \nu}=\eta_{\mu \nu}+\Phi k_\mu k_\nu,
\end{equation} 
where $\eta_{\mu\nu}$ is the flat metric, $\Phi$ is a scalar field and $k_\mu$ is a vector that is null and geodetic with respect to both $\eta_{\mu\nu}$ and  $g_{\mu \nu}$, i.e.
\begin{equation}\label{eq2.1.2}
    g^{\mu \nu}k_\mu k_\nu=\eta^{\mu \nu}k_\mu k_\nu=0, \qquad k^\mu \nabla_\mu k^\nu=k^\mu \partial_\mu k^\nu=0.
\end{equation}
The inverse of the spacetime metric is 
\begin{equation}\label{eq2.1.3}
    g^{\mu \nu}=\eta^{\mu \nu}-\Phi k^\mu k^\nu. 
\end{equation}
The index on $k_\mu$ can be raised with either $g_{\mu \nu}$ or $\eta_{\mu \nu}$, since
$g^{\mu \nu} k_\mu=\eta^{\mu \nu} k_\mu-\Phi (k^\mu k_\mu) k^\nu=\eta^{\mu \nu} k_\mu.
$

From the Kerr-Schild form metric \eqref{eq2.1.1}, the ``single copy'' Maxwell field is defined as
\begin{equation}\label{eq2.1.5}
    A_\mu=\Phi k_\mu.
\end{equation}
When $g_{\mu \nu}$ is a solution to Einstein's equations, the above Maxwell field is guaranteed to satisfy the Maxwell equations. $\Phi$ is the ``zeroth copy'' satisfying the Klein-Gordon equation on flat metric. 
The classical double copy of Kerr-Schild spacetimes thus maps certain classical solutions of Maxwell and Klein-Gordon equations to classical solutions of Einstein equations,
\begin{equation}\label{eq2.1.6}
    g_{\mu \nu}=\eta_{\mu \nu}+\frac{1}{\Phi}A_\mu A_\nu.
\end{equation}

It is also possible to extend the classical double copy to certain curved spacetime \cite{Carrillo-Gonzalez:2017iyj,Bahjat-Abbas:2017htu}. Consider the generalized Kerr-Schild metric
\begin{equation}\label{eq2.1.7}
    g_{\mu\nu}=\overline{g}_{\mu\nu}+\Phi k_\mu k_\nu,
\end{equation}
where the base metric $\overline{g}_{\mu\nu}$ now is curved. $k_\mu$ is again null and geodetic with respect to both the full and base metrics
\begin{equation}\label{eq2.1.8}
    g^{\mu \nu}k_\mu k_\nu=\overline{g}^{\mu \nu}k_\mu k_\nu=0, \qquad k^\mu \nabla_\mu k^\nu=k^\mu \overline{\nabla}_\mu k^\nu=0.
\end{equation}
The single and zeroth copy of \eqref{eq2.1.7} are still defined by $A_\mu=\Phi k_\mu$ and $\Phi$ which satisfy the Maxwell and Klein-Gordon equations on the base metric respectively.

\section{Local definition of cosmological horizons}
A co-dimension 2 spacelike surface $\mathcal{S}$ in four dimensional spacetime (with metric $g_{\mu\nu}$) always has two different future-directed null normal vectors, which are denoted by $n^\mu$ and $l^\mu$.
Taking the normalization $n_\mu l^\mu=-1$, we can write the induced metric on $\mathcal{S}$ as
\begin{equation}\label{eq2.3.1}
    h_{\mu \nu}=g_{\mu \nu}+n_\mu l_\nu+ l_\mu n_\nu.
\end{equation}
The expansion $\theta^{(V)}$ of $\mathcal{S}$ along a normal vector $V^\mu$ is defined as 
\begin{equation}\label{eq2.3.2}
    \theta^{(V)}=\mathcal{L}_V \ln \sqrt{\det h}=h^{\mu \nu}\nabla_\mu V_\nu,
\end{equation}
where $\mathcal{L}_V$ denotes the Lie derivative along $V^\mu$. 

Then, we introduce how to use the notion of expansion to define cosmological horizons in a Kerr-Schild spacetime\cite{Faraoni:2015ula,Chawla:2023bsu,Gourgoulhon:2008pu}. 
The method here is similar to that proposed to locate event horizons in \cite{Chawla:2023bsu}. 
In flat spacetime, we choose a hypersurface $\mathcal{H}$ with normal vector $\tilde{H}_\mu$. Then, we foliate $\mathcal{H}$ into co-dimension 2 spacelike surfaces $\mathcal{S}_\lambda$ parameterized by $\lambda$ such that Kerr-Schild vector $k_\mu$ is one of its two null normals. We also assume that $k_\mu$ is outgoing. The other 
ingoing null normal vector $l_\mu^{(0)}$ of $\mathcal{S}_\lambda$ can be determined as
\begin{equation}\label{eq3.1.4}
    l_\mu^{(0)}=\frac{H_\rho H_\sigma \eta^{\mu \nu}}{2}k_\mu+H_\mu,
\end{equation}
where $H_\mu=-\tilde{H}_\mu/k\cdot\tilde{H}$. The induced metric and its inverse of each $\mathcal{S}_\lambda$, inherited from $\eta_{\mu \nu}$, are given by
\bea\label{eq3.1.5}
\mathring{h}_{\mu \nu}=\eta_{\mu \nu}+ k_\mu l_\nu^{(0)}+l_\mu^{(0)}k_\nu,~\mathring{h}^{\mu \nu}=\eta^{\mu \nu}+k^\mu l_\rho^{(0)}\eta^{\nu\rho}+l_\rho^{(0)}\eta^{\rho \mu}k^\nu,
\eea
 where we use a circular accent to denote quantities defined on flat spacetime.

Now we move to the curved Kerr-Schild spacetime with metric $g_{\mu \nu}$ in the form of Eq.\eqref{eq2.1.1}.
Since the hyperspace $\mathcal{H}$ and its foliation $\mathcal{S}_\lambda$ are described in terms of the differential forms $H_\mu$ and $k_\mu$, they are in fact independent of the metric. Thus, the same hypersurface $\mathcal{H}$ and its foliation $\mathcal{S}_\lambda$ are used in the Kerr-Schild spacetime.        
Since $k_\mu$ is null with respect to both $\eta_{\mu \nu}$ and $g_{\mu\nu}$, it is still a null normal of the co-dimension 2 foliation $\mathcal{S}_\lambda$. The second null normal $l_\mu$ can be determined uniquely 
\begin{equation}\label{eq3.1.7}
    l_\mu=l_\mu^{(0)}-\frac{\Phi}{2}k_\mu.
\end{equation}
Then, the induced metric on $\mathcal{S}_\lambda$ and its inverse are
\begin{equation}
h_{\mu\nu}=g_{\mu\nu}+k_\mu l_\nu+l_\mu k_\nu, ~h^{\mu\nu}=g^{\mu\nu}+k^\mu l^\nu +l^\mu k^\nu.
\end{equation}

Now consider the expansion of $\mathcal{S}_\lambda$ along its null normals. In flat spacetime, using the induced metric in Eq.\eqref{eq3.1.5}, the expansions can be calculated as
\begin{equation}\label{eq3.2.1}
    \begin{split}
		\mathring{\theta}^{(k)}=\mathring{h}^{\mu\nu}\mathring{\nabla}_\mu k_\nu, \qquad \mathring{\theta}^{(l^{(0)})}=\mathring{h}^{\mu\nu}\mathring{\nabla}_\mu l_\nu^{(0)}.
    \end{split}
\end{equation}
Similarly, in Kerr-Schild spacetime, the expansions of $\mathcal{S}_\lambda$ are
\begin{equation}\label{3.3.2}
    \theta^{(k)}=h^{\mu\nu}\nabla_\mu k_\nu, \qquad \theta^{(l)}=h^{\mu\nu}\nabla_\mu l_\nu.
\end{equation}
In the appendix of \cite{Chawla:2023bsu}, the authors showed that for Kerr-Schild spacetimes the expansions on the $\eta_{\mu \nu}$ and $g_{\mu\nu}$ backgrounds are related via 
\begin{equation}\label{eq3.3.1}
    \theta^{(k)}=\mathring{\theta}^{(k)}, \qquad \theta^{(l)}=\mathring{\theta}^{(l^{(0)})}+\frac{\Phi}{2}\mathring{\theta}^{(k)}.
\end{equation}

In flat spacetime, the two expansions are always of opposite sign. For the assumption that $k_\mu$ is outgoing and $l^{(0)}_\mu$ is ingoing, $\mathring{\theta}^{(k)}>0$ and $\mathring{\theta}^{(l^{(0)})}<0$. $k_\mu$ describes a diverging null geodesic congruence and $l_\mu^{(0)}$ describes a converging null geodesic congruence.
In curved spacetime, the situation is different and a cosmological horizon is determined by the following criterion \cite{Faraoni:2015ula}
\bea\label{cret}
\theta^{(l)}=0,~\theta^{(k)}>0,~\mathcal{L}_k\theta^{(l)}>0.
\eea  
Due to the relations in Eq.\eqref{eq3.3.1}, the above criterion can be rewritten as
\bea
\mathring{\theta}^{(l^{(0)})}+\frac{\Phi}{2}\mathring{\theta}^{(k)}=0,~\mathring{\theta}^{(k)}>0,~\mathcal{L}_k(\mathring{\theta}^{(l^{(0)})}+\frac{\Phi}{2}\mathring{\theta}^{(k)})>0.
\eea
Thus, we locate the cosmological horizon of a curved spacetime by using just the corresponding single- and zeroth-copy quantities in base spacetime.

\section{Examples}
\label{sec:4}

\subsection{De Sitter spacetime}
\label{sub:4.1}

In static coordinates the four-dimensional de Sitter metric takes the form
\begin{equation}\label{eq4.1.1}
    ds^2=-(1-\frac{\Lambda r^2}{3})dt^2+(1-\frac{\Lambda r^2}{3})^{-1}dr^2+r^2d\theta^2+r^2 \sin^2 \theta d\varphi^2,
\end{equation}
where $\Lambda$ is the positive cosmological constant ($\Lambda>0$). Making a coordinate transformation $d\tau=dt \pm \frac{\frac{\Lambda r^2}{3}}{1-\frac{\Lambda r^2}{3}}dr$,  the de Sitter metric can be written in the Kerr-Schild form
\begin{equation}\label{eq4.1.2}
    ds^2=-d\tau^2+dr^2+r^2d\theta^2+r^2\sin^2 \theta d\varphi^2+\frac{\Lambda r^2}{3}(d\tau \pm dr)^2.
 \end{equation}
 The first four term in the above equation constitute the line element of flat metric $\eta_{\mu \nu}$ in spherical coordinates. The Kerr-Schild vector $k_\mu$ and the zeroth copy $\Phi$ are 
 \begin{equation}\label{eq4.1.3}
    k_\mu dx^\mu=-d\tau \mp dr, \qquad \Phi=\frac{\Lambda r^2}{3},
 \end{equation}
 from which the single copy gauge field constructed by $\Phi$ and $k_\mu$ is
  \begin{equation}\label{eq4.1.4}
    A_\mu dx^\mu=\Phi k_\mu dx^\mu= \frac{\Lambda r^2}{3}(-d\tau \mp dr),
 \end{equation}
which describes a uniform charge density filling all space by replacing cosmological constant $\lambda$ to charge density $\rho$ \cite{Luna:2015paa}. 
The Kerr-Schild vector's integral curves are future-directed and ingoing when taking a negative sign in Eq.\eqref{eq4.1.3}, and future-directed and outgoing when taking a positive sign. Here we take the positive sign, i.e. $k_\mu dx^\mu=-d\tau+dr$.  We now look at the family of hypersurfaces $\mathcal{H}_{(r)}$ at constant $r$ with normalized  normal vector
 \begin{equation}\label{eq4.1.5}
    H_\mu dx^\mu = -dr.
\end{equation}
On flat spacetime, the hypersurfaces $\mathcal{H}_{(r)}$ are foliated by co-dimension 2 spacelike surfaces $\mathcal{S}_\lambda$ normal to $k_\mu$. Then, according to Eq.\eqref{eq3.1.4}, the other null normal $l_\mu^{(0)}$ of $\mathcal{S}_\lambda$ is 
\bea\label{eq4.1.6}
        l_\mu^{(0)}dx^\mu=\frac{H_\rho H_\sigma \eta^{\rho \sigma}}{2}k_\mu dx^\mu + H_\mu dx^\mu=\frac{1}{2}(-d\tau-dr).
\eea
From the relation \eqref{eq3.1.7}, the ingoing normal of $\mathcal{S}_\lambda$ on de Sitter spacetime is
\bea\label{eq4.1.7}
l_\mu dx^\mu=l_\mu^{(0)}dx^\mu-\frac{\Phi}{2}k_\mu dx^\mu=-\frac{1}{2}(1-\frac{\Lambda r^2}{3})d\tau-\frac{1}{2}(1+\frac{\Lambda r^2}{r^2})dr.
\eea
Then, the expansions of  $\mathcal{S}_\lambda$ along the two null normals on the flat spacetime are
\begin{equation}\label{eq4.1.8}
    \mathring{\theta}^{(k)}=\frac{2}{r}, \qquad \mathring{\theta}^{(l^{(0)})}=-\frac{1}{r},
\end{equation}
and on the de Sitter spacetime are
\begin{equation}\label{eq4.1.9}
    \theta^{(k)}=\frac{2}{r}, \qquad \theta^{(l)}=\frac{\Lambda r^2-3}{3r}.
\end{equation}
The results in Eq.\eqref{eq4.1.8} is consistent with the general conclusion that $\mathring{\theta}^{(k)}$  and $\mathring{\theta}^{l^{(0)}}$ are of opposite sign in flat spacetime. From Eq.\eqref{eq4.1.9}, we see that in de Sitter spacetime $\theta^{(k)}$ is always positive while $\theta^{(l)}=0$ on the marginal surface $r=\sqrt{3/\Lambda}$. Then, we explicitly compute the Lie derivative of the expansion $\theta^{(l)}$ at $r=\sqrt{3/\Lambda}$,
\bea\label{eq4.1.10}   
        \mathcal{L}_k \theta^{(l)}\vert_{r=\sqrt{3/\Lambda}}=k^\mu\partial_\mu \theta^{(l)}|_{r=\sqrt{3/\Lambda}}=\frac{2\Lambda}{3}>0.
\eea
Thus, according to the criterion \eqref{cret}, the surface $r=\sqrt{3/\Lambda}$ is the cosmological horizon. Using just the information of a gauge field sourced by a uniform charge density filling all space on flat spacetime, we locate the cosmological horizon of the corresponding Kerr-Schild spacetime.   
        
The above discussion can be extended directly to $D$-dimensional de Sitter spacetime given in \cite{Lopez-Ortega:2009lik}, and in Kerr-Schild coordinates the line element is
\begin{equation}\label{eq4.1.11}
    ds^2=-d\tau^2+dr^2+r^2d\Omega_{d-2}^2+\frac{ r^2}{L^2}(d\tau \pm dr)^2,
\end{equation} 
where $L^2=\frac{(D-1)(D-2)}{2 \Lambda} (D>2)$ and $\Lambda$ is the cosmological constant. The zeroth copy and Kerr-Schild vector are
\begin{equation}\label{eq4.1.12}
    \Phi=\frac{r^2}{L^2}, \qquad k_\mu dx^\mu=-d\tau \mp dr.
\end{equation}
Similarly, we take $k_\mu=(-1,1,0,...,0)$ as the outgoing null normal and the normalized one-form is $H_\mu dx^\mu=-dr$.
The other null normal $l_\mu^{(0)}$ of the foliation $\mathcal{S}_\lambda$ is
\begin{equation}\label{eq4.1.13}
    \begin{split}
        l_\mu^{(0)} dx^\mu=(\frac{H_\rho H_\sigma \eta^{\rho \sigma}}{2}k_\mu  + H_\mu)dx^\mu=\frac{1}{2}(-d\tau-dr).
    \end{split}
\end{equation}
On the $D$-dimensional de Sitter spacetime, the corresponding null normal is
\begin{equation}\label{eq4.1.14}
    l_\mu dx^\mu = \frac{1}{2}(-1+\frac{r^2}{L^2})d\tau-\frac{1}{2}(1+\frac{r^2}{L^2})dr.
\end{equation}
 Then, the expansions of the foliation along the two null normals $k^\mu$ and $l^\mu$ on Kerr-Schild spacetime can be obtained,
\begin{equation}\label{eq4.1.16}
    \theta^{(k)}=\frac{D-2}{r}, \qquad \theta^{(l)}=\frac{D-2}{2}(-\frac{1}{r}+\frac{r}{L^2}).
\end{equation}
We can see $\theta^{(l)}$ becomes zero when $r=L=\sqrt{\frac{(D-1)(D-2)}{2 \Lambda}}$. Then, we also check that
\begin{equation}\label{eq4.1.17}
    \begin{split}
        \mathcal{L}_k \theta^{(l)}|_{r=L}=k^\mu\partial_\mu \theta^{(l)}|_{r=L}
        =\frac{D-2}{L^2}>0.
    \end{split}
\end{equation}
Thus, we locate the cosmological horizon at $r=\sqrt{\frac{(D-1)(D-2)}{2 \Lambda}}$.

\subsection{De Sitter black holes}
\label{sub:4.2}
The line element of a four-dimensional Schwarzschild-de Sitter black hole is
\begin{equation}\label{eq4.2.1}
    ds^2=-(1-\frac{2M}{r}-\frac{\Lambda r^2}{3})dt^2+(1-\frac{2M}{r}-\frac{\Lambda r^2}{3})^{-1}dr^2+r^2d\theta^2+r^2\sin^2\theta d\varphi^2,
\end{equation}
where $M$ is the mass parameter and $\Lambda$ is the positive cosmological constant, and its Kerr-Schild form is
\begin{equation}\label{eq4.2.2}
    \begin{split}
        ds^2=-d\tau^2+dr^2+r^2d\theta^2+r^2\sin^2\theta d\varphi^2+(\frac{2M}{r}+\frac{\Lambda r^2}{3})(d\tau \pm dr)^2 
    \end{split}
\end{equation}
where the zeroth copy and Kerr-Schild vector are
\begin{equation}\label{eq4.2.3}
    \Phi=\frac{2M}{r}+\frac{\Lambda r^2}{3}, \qquad k_\mu dx^\mu=-d\tau\mp dr.
\end{equation}
The corresponding single copy is defined as 
$A_\mu dx^\mu=\Phi k_\mu dx^\mu= (\frac{2M}{r}+\frac{\Lambda r^2}{3})(-d\tau \mp dr)$,
which is the gauge field sourced by a point charge plus a uniform charge density filling all space. 
        
We choose $k_\mu dx^\mu=-d\tau+dr$ as the outgoing null normal. Now, we consider the family of hypersurfaces $\mathcal{H}_{(r)}$ at constant $r$ with normalized normal vector $H_\mu dx^\mu=-dr$. The other ingoing null normal on flat spacetime is $l_\mu^{(0)}dx^\mu=\frac{1}{2}(-d\tau-dr)$. On Kerr-Schild spacetime, the ingoing null normal is 
\begin{equation}\label{eq4.2.5}
    l_\mu dx^\mu=-\frac{1}{2}(1-\frac{2M}{r}-\frac{\Lambda r^2}{3})d\tau+\frac{1}{2}(1+\frac{2M}{r}+\frac{\Lambda r^2}{3})dr.
\end{equation}
The expansions in flat spacetime are
$
    \mathring{\theta}^{(k)}=\frac{2}{r}, \mathring{\theta}^{(l^{(0)})}=-\frac{1}{r},
$
and in the Schwarzschild-de Sitter spacetime are
\begin{equation}\label{eq4.2.7}
    \theta^{(k)}=\frac{2}{r}, \qquad \theta^{(l)}=\frac{f(r)}{3r^2},
\end{equation}
where $f(r)=\Lambda r^3-3r+6M$. 

The Schwarzschild-de Sitter black hole has an black hole event horizon and a cosmological horizon when $0<M<1/3\sqrt{\Lambda}$ is satisfied. The locations of these horizons correspond to the positive roots of $1-\frac{2M}{r}-\frac{\Lambda r^2}{3}=0$, i.e. $f(r)=\Lambda r^3-3r+6M=0$. Equation $f(r)=0$ has two positive roots $r_1$, $r_2$ and a negative root $r_0$ which satisfy $r_0<r_1<r_2$ and $r_0=-(r_1+r_2)$ \cite{Fernandes:2019ige}. Thus, $f(r)$ can be written as $f(r)=(r-r_0)(r-r_1)(r-r_2)$.
We compute the Lie derivative of the expansion $\theta^{(l)}$ along $k^\mu$ at $r=r_2$ and obtain
\begin{equation}\label{eq4.2.9}
    \begin{split}
        \mathcal{L}_k \theta^{(l)}\vert_{r=r_2}=k^\mu\partial_\mu \theta^{(l)}|_{r=r_2}
       =\frac{r_2(r_2^2-r_1^2)+r_2^2(r_2-r_1)}{3r_2^2}>0.
    \end{split}
\end{equation} 
Thus, the surface $r=r_2$ is the cosmological horizon. One can check that the above derivative is negative at $r=r_1$.
        
Then, we consider the four-dimensional Reissner-Nordstrom-de Sitter (RNdS) black hole with a Kerr-Schild form as follows
\begin{equation}\label{eq4.2.10}
    \begin{split}
        ds^2=-d\tau^2+dr^2+r^2d\theta^2+r^2\sin^2\theta d\varphi^2+(\frac{2M}{r}-\frac{Q^2}{r^2}+\frac{\Lambda r^2}{3})(d\tau \pm dr)^2,
    \end{split}
\end{equation}
where the zeroth copy and Kerr-Schild vector are
\begin{equation}\label{eq4.2.11}
    \Phi=\frac{2M}{r}-\frac{Q^2}{r^2}+\frac{\Lambda r^2}{3}, \qquad k_\mu dx^\mu=-d\tau\mp dr.
\end{equation} 
We also take $k_\mu dx^\mu=-d\tau+dr$ and consider the family of hypersurfaces $\mathcal{H}_{(r)}$ at constant $r$ with normalized normal vector $H_\mu dx^\mu=-dr$.
The ingoing null normal in flat spacetime is also $l_\mu^{(0)}dx^\mu=\frac{1}{2}(-d\tau-dr)$ and in Kerr-Schild spacetime is
\begin{equation}\label{eq4.2.12}
    l_\mu dx^\mu=-\frac{1}{2}(1-\frac{2M}{r}-\frac{\Lambda r^2}{3})d\tau+\frac{1}{2}(1+\frac{2M}{r}+\frac{\Lambda r^2}{3})dr.
\end{equation}
The expansions in flat spacetime are
$\mathring{\theta}^{(k)}=\frac{2}{r}, \mathring{\theta}^{(l^{(0)})}=-\frac{1}{r}$
and in the RNdS spacetime are
\begin{equation}\label{eq4.2.14}
    \theta^{(k)}=\frac{2}{r}, \qquad \theta^{(l)}=\frac{g(r)}{3r^2},
\end{equation}
where $g(r)=\Lambda r^4-3r^2+6Mr-3Q^2$. 
Equation $g(r)=0$ has three positive roots $R_1, R_2, R_3$ and a negative root $R_0$, which satisfy $R_0<R_1<R_2<R_3$ and $R_0=-(R_1+R_2+R_3)$ \cite{Chambers:1997ef}. We rewrite the function $g(r)$ as $g(r)= (r-R_0)(r-R_1)(r-R_2)(r-R3)$ and
 compute the Lie derivative of the expansion $\theta^{(l)}$ along $k^\mu$ at $r=R_3$,
\begin{equation}\label{eq4.2.16}
    \begin{split}
        \mathcal{L}_k \theta^{(l)}\vert_{r=R_3}=k^\mu\partial_\mu \theta^{(l)}|_{r=R_3}
     =\frac{(R_3-R_1)(R_3-R_2)(R_1+R_2+2R_3)}{3R_3^2}>0.
    \end{split}
\end{equation} 
Thus, we locate the surface $r=R_3$ as the cosmological horizon.

For higher-dimensional Schwarzschild(RN)-de Sitter spacetime, their cosmological horizons can be located similarly with the double copy method.        
In addition, we point out that if we choose the ingoing Kerr-Schild vector as one null normal of the foliation, we can also locate the event horizons of the de Sitter black holes with the criterions proposed in \cite{Chawla:2023bsu,Faraoni:2015ula}.

\subsection{Four-dimensional black plane}
\label{sub:4.4}

In this subsection we consider the case of four-dimensional black plane with zero charge. The Kerr-Schild double copy about black plane spacetime was studied in \cite{Carrillo-Gonzalez:2017iyj}, where the background metric is a plane de Sitter spacetime. 
Although the background  spacetime is curved, every step of the procedure for locating cosmological horizon still holds. Next, we locate the cosmological horizon using single and zeroth copy data.         
The black plane metric with vanishing charge is given by \cite{Cai:1996eg}
\begin{equation}\label{eq4.4.1}
    ds^2=-(\alpha^2 r^2-\frac{4\pi M}{\alpha^2 r})dt^2+(\alpha^2 r^2-\frac{4\pi M}{\alpha^2 r})^{-1}dr^2+\alpha^2 r^2(dx^2+dy^2),
\end{equation}
where $M$ is the ADM mass and $3\alpha^2=-\Lambda<0$. This solution is asymptotically de Sitter not only in the transverse directions, but also in the membrane directions and has a single cosmological horizon at $r=(\frac{4\pi M}{\alpha^4})^{\frac{1}{3}}$. The vacuum solution ($M=0$) corresponding to \eqref{eq4.4.1} is
\begin{equation}\label{eq4.4.2}
    ds^2=-\alpha^2 r^2 dt^2+(\alpha^2 r^2)^{-1}dr^2+\alpha^2 r^2(dx^2+dy^2),
\end{equation}
which is plane de Sitter spacetime. Regard this spacetime as a background metric, the black plane metric in Kerr-Schild coordinates is 
\begin{equation}\label{eq4.4.3}
    \begin{split}
        ds^2=-\alpha^2 r^2 d\tau^2+(\alpha^2 r^2)^{-1}dr^2+\alpha^2 r^2(dx^2+dy^2)
        +(\alpha^2 r^2 \frac{4\pi M}{\alpha^4 r^3})(d\tau \pm \frac{1}{\alpha^2 r^2}dr)^2,
    \end{split}
\end{equation}
where the zeroth copy and Kerr-Schild vector are 
\begin{equation}\label{eq4.4.4}
    \Phi= \alpha^2 r^2 \frac{4\pi M}{\alpha^4 r^3}, \qquad k_\mu dx^\mu=-d\tau \mp \frac{1}{\alpha^2 r^2}dr.
\end{equation}
        
We take the outgoing Kerr-Schild vector $k_\mu dx^\mu=-d\tau + \frac{1}{\alpha^2 r^2}dr$ in plane de Sitter spacetime. We also consider the family of hypersurfaces $\mathcal{H}_{(r)}$ at constant r with normalized normal $H_\mu dx^\mu=-dr$. The ingoing null normal in the plane de Sitter spacetime is $l_\mu^{(0)}dx^\mu=-\frac{1}{2}\alpha^2 r^2d\tau-\frac{1}{2}dr$. In the Kerr-Schild spacetime, the ingoing normal is
\begin{equation}\label{eq4.4.5}
    l_\mu dx^\mu=-(\frac{2\pi M}{\alpha^2 r}-\frac{\alpha^2 r^2}{2})d\tau+(\frac{2\pi M}{\alpha^4 r^3}+\frac{1}{2})dr.
\end{equation}
The expansions in the plane de Sitter spacetime are
$\bar{\theta}^{(k)}=\frac{2}{r}, \bar{\theta}^{(l^{(0)})}=-\alpha^2 r.
$
Please note that $k_\mu$ and $l_\mu^{(0)}$ have the same sign due to the background is plane de Sitter spacetime, which has a cosmological singularity at $r=0$.
The expansions in black plane spacetime are
\begin{equation}\label{eq4.4.7}
    \theta^{(k)}=\frac{2}{r}, \qquad \theta^{(l)}=\frac{4\pi M-\alpha^4 r^3}{\alpha^2r^2}.
\end{equation}
 $\theta^{(l)}=0$ at $r=(\frac{4\pi M}{\alpha^4})^{\frac{1}{3}}$ . The Lie derivative of the expansion $\theta^{(l)}$ at $r=(\frac{4\pi M}{\alpha^4})^{\frac{1}{3}}$ is
\begin{equation}\label{eq4.4.8}
    \begin{split}
        \mathcal{L}_k \theta^{(l)}\vert_{r=(\frac{4\pi M}{\alpha^4})^{\frac{1}{3}}}=k^\mu\partial_\mu \theta^{(l)}|_{r=(\frac{4\pi M}{\alpha^4})^{\frac{1}{3}}}
     =-3\alpha^2>0.
    \end{split}
\end{equation} 
Thus, the surface $r=(\frac{4\pi M}{\alpha^4})^{\frac{1}{3}}$ is the cosmological horizon.

\section{Summary}
\label{sec:5}

In this paper, we show through several examples how to locate the cosmological horizons for stationary Kerr-Schild spacetime in the classical double copy paradigm. Our work is a generalization of that on the locating of black hole event horizons in \cite{Chawla:2023bsu}. It is found that the cosmological horizons of the de Sitter spacetime, Schwarzschild(RN)-de Sitter black holes and black plane spacetime indeed can be located by only using the corresponding single- and zeroth-copy data. 

When foliating the chosen hypersurfaces to co-dimension 2 spacelike surfaces $\mathcal{S}_\l$, we choose an outgoing Kerr-Schild vector $k_\mu$ as a null normal of the foliation, instead of an ingoing one in the locating of black hole event horizons \cite{Chawla:2023bsu}. An independent ingoing null normal of $\mathcal{S}_\l$ in the flat base metric can be determined in Eq.\eqref{eq3.1.4}. After computing the expansions of  $\mathcal{S}_\l$ along the two null normals, the location of cosmological horizon is determined by Eqs.\eqref{eq3.3.1}\eqref{cret}. 

For the de Sitter spacetime and Schwarzschild(RN)-de Sitter black hole cases, the base metrics are all flat. Last but maybe more interesting is the black plane case, where 
the base metric is a curved plane de Sitter spacetime. Our result shows that the method for locating cosmological horizon still holds in this curved background case. 

The discussed asymptotically de Sitter black holes in this paper are all spherically symmetric. It will be interesting to consider the application of the double copy method to the rotating or axially symmetric spacetime with cosmological horizons.

\section*{Acknowledgments}\label{Ack}
This work is supported in part by Guangdong Major Project of Basic and Applied Basic Research (No.2020B0301030008).

\end{document}